\documentclass[pre,aps,showpacs]{revtex4}

\usepackage{amssymb}
\usepackage{graphicx}

\begin{document}

\title
{Mode I fracture in a nonlinear lattice with
viscoelastic forces}
\author
{Shay I. Heizler and David A. Kessler}
\affiliation{Department of Physics, Bar-Ilan University,
 Ramat-Gan, ISRAEL}
\author
{Herbert Levine}
\affiliation{Department of Physics, Univ. of Cal., San Diego, La
Jolla, CA 92093-0319 USA}
\begin{abstract}
We study Mode I fracture in a viscoelastic lattice model with a
nonlinear force law, with a focus on the velocity and linear
stability of the steady-state propagating solution. This study is
a continuation both of the study of the piece-wise linear model in
Mode I, and the study of more general nonlinear force laws in Mode
III fracture. At small driving, there is a strong dependency of
the velocity curve on the dissipation and a strong sensitivity to
the smoothness of the force law at large dissipation. At large
driving we calculate, via a linear stability analysis, the
critical velocity for the onset of instability as a function of
the smoothness, the dissipation and the ratio of lattice spacing
to critical extension. This critical velocity is seen to be very
sensitive to these parameters. We confirm our calculations via
direct numerical simulations of the initial value problem.
\end{abstract}
\maketitle
\section{INTRODUCTION}
The problem of extensional (Mode I) fracture has received
increasing attention in the physics community in the last
decade~\cite{review}.  The experiments in amorphous materials done
by Fineberg, et al.~\cite{fineberg} and in ordered materials by
Hauch, et al.~\cite{silicon}, showing interesting dynamical
behavior for large velocity cracks have been at the center of this
growing interest. From a theoretical point of view, the
singularities at the crack tip associated with continuum
treatments of the crack make the problem challenging. The presence
of these singularities necessitates a treatment of the crack tip
at the microscopic level, the singularities being regularized by
the small scale dynamics. One line of attack on this problem,
initiated by Slepyan~\cite{slepyan}, has been through the study of
lattice models of cracks. These lattice models are simpler than
full atomistic simulations~\cite{mdsim} in that the connectivity
of the atoms is specified from the beginning, and so dislocations
are excluded.

Much progress has been made in understanding steady-state
propagation of cracks in lattice systems using an ideally-brittle
piece-wise linear force
law~\cite{slepyan,slepyan2,mg,kl1,crack3,pechenik,newslepyan,gerde}.
Here the particles interact with Hookean springs, which break at
some critical extension, after which they exert no force. With
these piece-wise-linear interactions, the model admits an analytic
solution via the Wiener-Hopf technique.  This solution has been
carried out both for Mode III and Mode I cracks, for both finite
width and infinitely wide systems, with and without dissipation
(Stokes or Kelvin type). There have also been some recent results
on mode II cracks, with possible relevance for an understanding of
friction~\cite{gerde}

The piece-wise linear model, despite its analytic simplicity,
exhibits some undesirable features~\cite{pechenik}. For small
dissipation, the solutions are inconsistent at small velocities,
in that bonds on the crack surface which are assumed to crack at
time some specific time in fact are seen to reach the critical
displacement and hence to crack earlier. At large velocities, the
analytic solutions are again inconsistent, this time due to the
breaking of additional bonds off the nominal crack surface. Beyond
this point, the analytic methods are unable to tell us anything
about the true dynamics of the system. An additional limitation of
the piece-wise-linear force law is that it greatly complicates the
task of constructing a linear stability treatment of the
steady-state crack. This is due of course to the discontinuous
nature of the force law.

For the case of Mode III fracture, Kessler and
Levine~\cite{nonlinear} studied a lattice model with a family of
continuous non-linear force laws labeled by a tunable smoothness
parameter $\alpha$. Such continuous forces should presumably more
closely model the actual physical situation. With this lattice
model, they investigated both arrested cracks~\cite{arrested} and
propagating cracks~\cite{nonlinear}. Consistent steady-state
solutions were found at all driving; the previously discovered
inconsistent solutions at large driving of the piece-wise linear
limit were converted to consistent but linearly unstable solutions
for the continuous force law.

In this paper we will study Mode I fracture with a suitable
adaptation to vector forces of this class of continuous force law.
We shall study steady-state solutions of the model both for small
and large driving, and their dependence on the dissipation, the
critical bond extension (relative to the lattice constant) and
smoothness. Our results will be compared those of both the
piece-wise linear limit Mode I calculations~\cite{pechenik} and
the continuous force Mode III analysis~\cite{nonlinear}. We shall
also study the instabilities of the theory at large driving,
finding a number of transitions between dominant modes as a
function of dissipation; these findings will be confirmed by
direct numerical simulations. We will also briefly consider the
nature of the dynamics, post-instability. Finally, we shall draw
some conclusions and discuss directions for further research.

\section{MODEL AND GENERAL METHODOLOGY}

In this paper we study extensional (Mode I) cracks in a nonlinear,
two dimension hexagonal lattice model. Mass points located at the
lattice sites are coupled to their six nearest-neighbors by
nonlinear, viscoelastic central forces. The force that particle
$2$ exerts on particle $1$ is taken to be:
\begin{equation}
\vec{f}_{1,2}=(r_{1,2}-a)\ \frac{1+\tanh [\alpha (\varepsilon -r_{1,2})]}{%
1+\tanh (\alpha )}\ \hat{r}_{1,2}
\end{equation}
where $a$ is the lattice scale,
$\vec{r}_{1,2}=\vec{x}_{1}-\vec{x}_{2}$, $r_{1,2}=|\vec{r}_{1,2}|$
is the distance between the atoms and
$\hat{r}_{1,2}=\frac{\vec{r}_{1,2}}{r_{1,2}}$. The threshold
$\varepsilon $ is the distance between the atoms at which the
spring can be considered to break. $\alpha $ is a parameter that
determines the smoothness of the potential. In the limit $\alpha
\to \infty $ the force becomes perfectly brittle, dropping
immediately to $0$ as $r_{1,2}$ exceeds $\varepsilon$. Decreasing
$\alpha$ smoothes the force, so that it decays away over a
distance $1/\alpha$. Now the bonds never crack totally, there is
always some (exponentially small) force, even at large $r_{1,2}$.
Note that, compared with the similar force law studied at Mode
III~\cite{nonlinear}, this model has an additional parameter,
namely $a/\varepsilon$ the ratio of the lattice constant to the
breaking extension; this parameter governs the extent to which the
force direction changes as the displacement grows. Only in the
joint limit $\alpha\to\infty$, $a/\varepsilon \to \infty$ do we recover the 
piece-wise
linear force law studied in Mode I by Kulamekhtova, Saraikin and
Slepyan~\cite{slepyan2}, Marder and Liu~\cite{marderliu} and
Pechenik, Kessler and Levine~\cite{pechenik}. In the following we
choose our length scale such that $\varepsilon =1$.

In addition to the purely conservative force, we introduce a
Kelvin-type viscosity with a viscosity parameter $\eta$ with the
force law~\cite{kl1,crack3,pechenik,langer}:
\begin{equation}
\vec{g}_{1,2}=\eta k_{1,2}(\vec{v}_{1,2}\cdot
\hat{r}_{1,2})\hat{r}_{1,2}
\end{equation}
where $\vec{v}_{1,2}=\vec{v}_{1}-\vec{v}_{2}$ and $k_{1,2}$ is an
effective spring constant:
\begin{equation}
k_{1,2}=f_{1,2}/(r_{1,2}-a).
\end{equation}
These forces define our model. We can now write the equation of
motion for the displacement $u$ of a mass point away from its
lattice position:
\begin{equation}
\label{eom}
\frac{d^2 \vec u (\vec{x})}{dt^2}=%
\mathop{\displaystyle\sum}%
\limits_{\bar{x}^{\prime }\in nn}(\vec{f}_{\vec{x},\vec{x}^{\prime }}+\vec{g}%
_{\vec{x},\vec{x}^{\prime }}).
\end{equation}

We work on a hexagonal lattice with $2N+2$ rows in the $y$
direction, indexed by $j=-N\ldots N+1$. The rows are separated by
a distance $\sqrt{3}/2\cdot a$, so that $y_j=(2j-1)a\sqrt{3}/4$.
We apply a constant displacement $\vec u=\pm \Delta \hat y$ to the
top- and bottom-most
 rows. The (metastable) uncracked state is that of uniform strain
\begin{equation}
\vec u^{U}(x,y_j)= \frac{2j-1}{2N+1}\Delta \;\hat{y}.
\end{equation}
In the equilibrium cracked state, the upper half of the rows have
$\vec u=+\Delta \hat y$,
and the lower half $\vec u=-\Delta\hat y $.

Initially, we will be interested in the case of steady state
cracks where the displacement has the Slepyan form:
$\vec{u}(t,x,y)=\vec{u}(t-x/v,y)$ where $x,y$ label the position
of a particle on the lattice, and $v$ is the crack propagation
speed~\cite{slepyan,slepyan2}. This ansatz reduces the problem to
one of solving for the time development of the displacement for
just $2N$ particles, i.e. one particle on each of the
unconstrained rows. In addition, we focus on symmetric cracks so
we need consider just one side of the lattice, imposing the
symmetry condition:\ $%
u_{y}(t,y)=-u_{y,}(t-a/(2v),-y)$ and $u_{x}(t,y)=u_{x}(t-a/(2v),-y)$.

As opposed to the previous analyses done in the piece-wise linear
limit ($\alpha,a\to \infty$)~\cite{pechenik}, here with finite
$\alpha ,a$ we must resort to a numerical procedure. Specifically,
we discretize time using a small but finite $dt$, and further
limit $-T\leqslant t\leqslant T$. This reduces the problem to a
set of $2N(2T+1$) coupled equations. Since the equations of motion
are dependent only on nearest neighbors, the system has a banded
structure. The equations of motions are nonlinear so we use
Newton's method to solve the problem, using standard subroutines
for solving banded matrices to find the Newton update. Now, for
each given $v$ of the crack, we want to find the driving
displacement $\Delta $. We have $2N(2T+1)+1$ unknowns (the $N$
vector functions for the entire time range plus $\Delta $%
); this is one more than the number of equations, reflecting the
time-translation invariance of the sought-after solution. Thus we
add an additional equation fixing $u_{y=\sqrt{3/4}}(0)=\textit{const}$
to lift the degeneracy. Let us note that it can be shown
that the discretized problem has
one more linear mode that diverges as $x\to -\infty $ than diverges at $%
x\to \infty .$ To treat this, we enforce the equation of motion for $%
-T-dt\leqslant t\leqslant T-dt$ whereas the variable run from $-T\leqslant
t\leqslant T$~\cite{kl1}. The system then has $2N(n_{b}+3)-2$ bands
below the diagonal and $2N(n_{b}-1)+3$ upper bands, where $n_{b}$ is a even
number with $v=a/(n_{b}dt).$

The only remaining technical issue is that the system as presented
does not have a completely banded structure due to the fact that
many equations depend on $\Delta $. This problem can be
circumvented using the algorithm of Kessler and Levine (see the
appendix of Ref.~\cite{nonlinear}).

\section{THE SMALL VELOCITY REGIME}

It is reasonable to expect that smoothing out the potential can
have a large effect in the small velocity regime. In particular,
the fact that solutions in the piece-wise linear limit become
inconsistent at small enough velocity means that making $\alpha$
finite must make a significant difference. For mode III
~(\cite{nonlinear}), decreasing $\alpha$ eliminated the
oscillations in the  $v(\Delta) $ curve, in accord with this
reasoning; but, the oscillations were already not present at large
enough $\eta$ and thus changing $\alpha $ has almost no effect on
those steady-state curves.
\begin{figure}
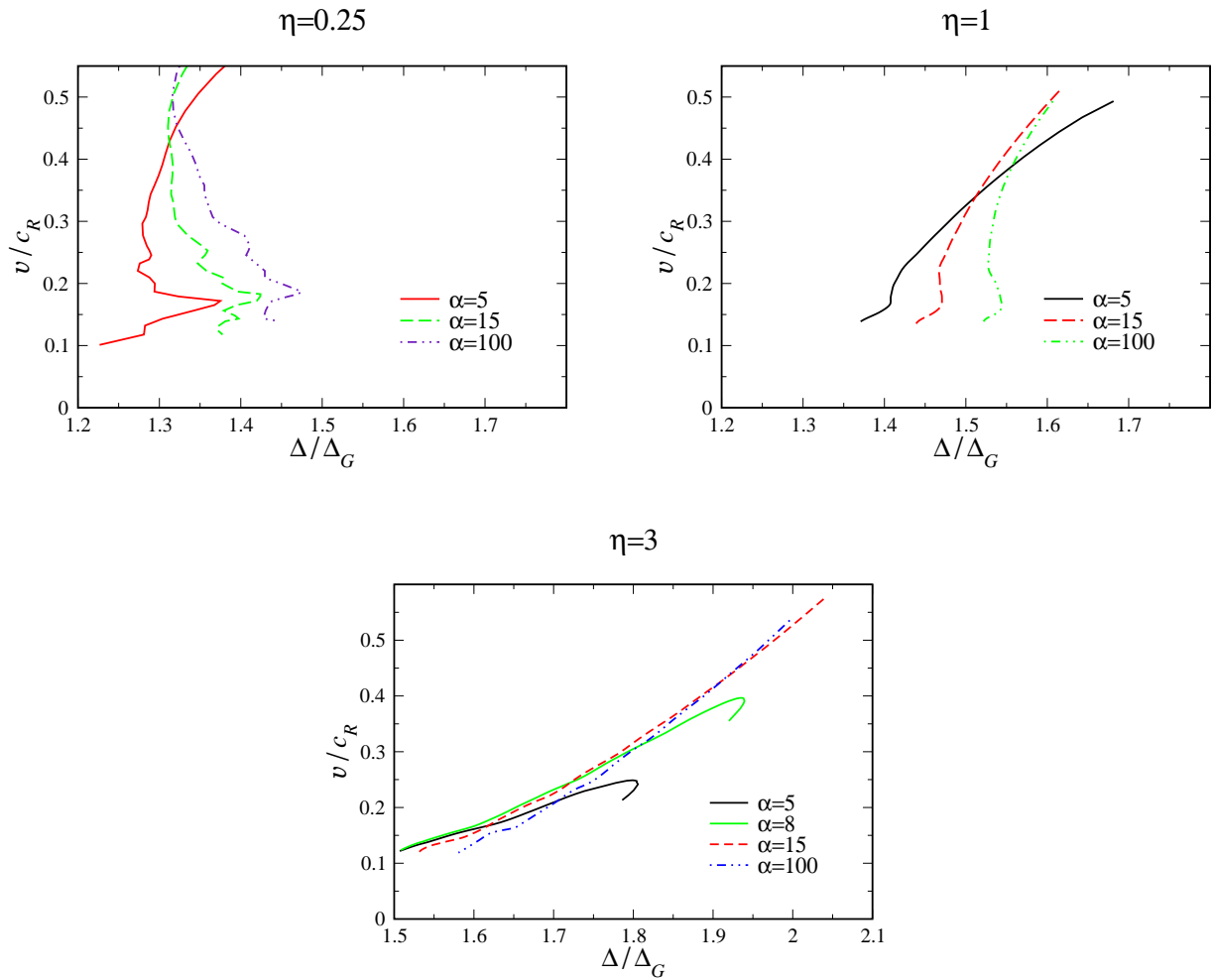

\centerline{
\includegraphics*[width=7.5cm]{1.eps}
\hskip 0.4in
\includegraphics*[width=7.5cm]{2.eps}}
\vskip 0.3in
\centerline{\includegraphics*[width=7.5cm]{3.eps}}
\caption{(a-c) Dependence of $v/c_{R}$ on $\Delta/\Delta_{G}$ for $\alpha=
 5,\ 15$ and $100$, for $\eta =0.25,\ 1$ and $3$. $N=3$, $dt=0.1$ and $a=4$. In (c) we
 present data also for $\alpha =8$.
}
\label{fig1}
\end{figure}

Here for Mode I we find a somewhat different picture. First we
present in Fig. \ref{fig1}(a-c) a graph of $v$ versus $\Delta $
for the steady state solution. $v$ is normalized with respect to
the Rayleigh
surface wave speed, the limiting crack speed (in the $%
N\to \infty $ limit) as the elastic field has to propagate along
the crack surface:\ $C_{R}=\frac{\sqrt{3-\sqrt{3}}}{2}a\simeq
0.563 \ a$. $\Delta $ is normalized by the Griffith value, $\Delta
_{G}$, the value for which fracture is energetically allowed. This
value is a function of $\alpha ,a$ and of course $N$. Data is
presented for $N=3$ because computing the very low velocity
solutions is very hard at large $N$, due to the bandwidth which
scale as $N/v$. We do not expect the picture at larger $N$ to be
qualitatively different~\cite{pechenik}.

At small $\eta $ the curves have oscillations for all  $%
\alpha $, unlike the Mode III case, where oscillations were present
only for large $\alpha $. At large $\alpha $, in both cases the
graphs are qualitatively similar to those obtained directly in the
piece-wise linear limit(PLL), though the large $\alpha$ curves do
not actually converge to those curves. This is due to the
aforementioned fact that the solutions of the PLL are actually
inconsistent for small velocity whereas the solutions of the
finite $\alpha$ model are always
well-defined~\cite{nonlinear,pechenik,mg}. As we increase $\eta $
the oscillations disappear and the curves become smoother. This
behavior, as can be seen in the $\eta =1\,\,$\ data (Fig.
\ref{fig1}(b)), happens for all $\alpha $, but is especially
pronounced for the smaller $\alpha $'s as expected. These trends are
also seen in the Mode III case~\cite{nonlinear}. As we increase
$\eta$ to still higher values, for example $\eta =3$, (Fig.
\ref{fig1}(c)), a maximal velocity can be seen at small $\alpha$.
The existence of a maximal velocity in steady-state cracks is a
general feature, and will be discussed at length in the next
section on the large velocity regime. Here we just note that for
large $\eta$ and small $\alpha$, the maximal velocity can be quite
low, e.g. $v_{max}\simeq 0.25C_{R}$ at $\alpha =5$, $a=4$, $\eta
=3$, and decreases for decreasing $\alpha$. This behavior has no
parallel in the Mode III case. At small $\Delta$, decreasing
$\alpha$ increases the velocity,  but for large $\Delta$ the
situation is reversed, with the curves with the smaller $\alpha $
having lower velocity.  This is a result of the fact that in our
nonlinear potential the bonds never really break and continue to
impede the crack's progress. On the other hand, at small $\Delta$,
small $\alpha$ weakens the lattice trapping effect, resulting in
larger velocity than for larger $\alpha$.

\begin{figure}
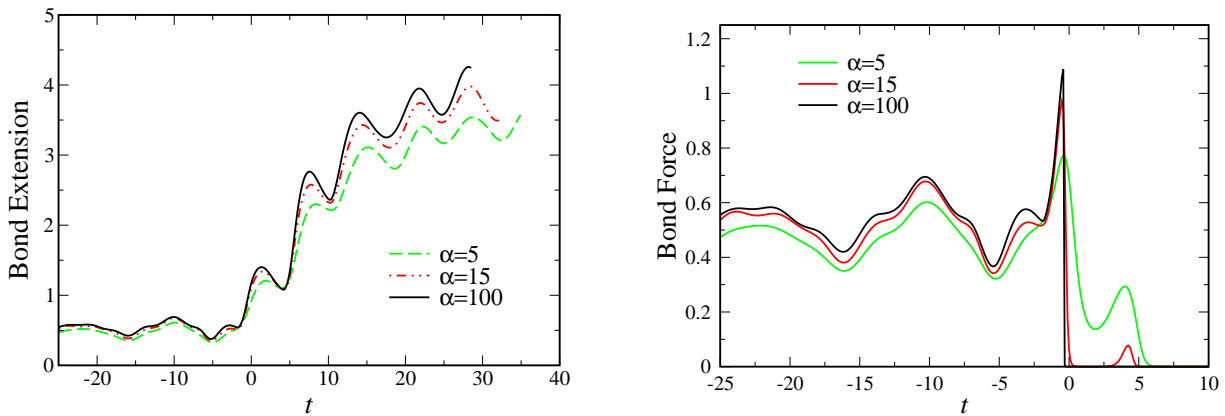

\centerline{\includegraphics*[width=7.5cm]{4.eps}
\hskip 0.4in
\includegraphics*[width=7.5cm]{5.eps}}
\caption{Bond extension (a) and force (b) along crack surface versus $t$, for $\alpha =
5,\ 15$ and $100$. In both figures, $N=3,\ \eta=0.25,\ v=0.148$ and $dt=0.1033$.
}
\label{fig2}
\end{figure}

We still have to address the question of why the curves oscillates
at low $\alpha$ for small $\eta$, in contradistinction to the Mode
III case~\cite{nonlinear}. It is interesting to consider (see Fig.
\ref{fig2}) the bond extension and the bond force for a
representative small velocity case, $v=0.148C_{R}$. In the
piece-wise linear case at small $\eta $ for both Mode I and Mode
III the underdamping of the backward running waves leads to
pre-cracking. This means that the bond extension rises above unity
($\varepsilon =1)$ before $t=0$ and then comes back to unity
before cracking once and for all~\cite{kl1,mg}. For finite $\alpha
$, while there cannot be any pre-cracking (we don't have to
postulate in advance when a bond will break), there is a remnant
of this tendency. After cracking, the bond extension first rises
and then falls to a value close to the critical extension before
rising again and then permanently leaving the region close to
critical extension. This force oscillation is ultimately what is
responsible for the oscillations in the $v(\Delta) $ plot. This
situation is very different from that in Mode III, where for
finite $\alpha $, the bond extension increases monotonically
through the critical region.

\begin{figure}
\centering{
\vskip 0.3in
\includegraphics*[width=8cm]{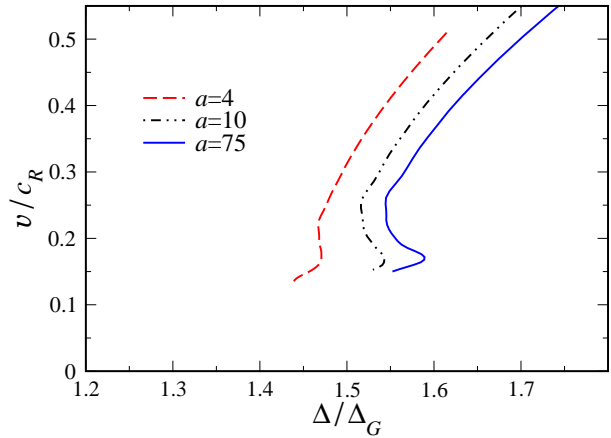}}
\caption{ Dependence of $v/c_{R}$ on $\Delta/\Delta_{G}$ for $a=
 4,\ 10$ and $75$, for $\eta =1$. $N=3$, $dt=0.13$ and $\alpha =15$.
}
\label{figext}
\end{figure}

We checked also the effect of the lattice scale constant $a$ on the curves for
the small velocity regime at $\eta =1$, $\alpha =15$. We present the data on
Fig. \ref{figext}.
We see that there is no dramatic effect of the lattice scale
constant in the small velocity regime. There is a backward branch
for all $a$ at about the same velocity, with the effect
accentuated at large $a$.

\section{LARGE VELOCITY REGIME}

In the large velocity regime, it is again clear that smoothing the
potential must make significant modifications to the steady state
solutions, as it is known that in the piece-wise linear model the
solution is inconsistent due to the breaking of bonds off the
assumed crack surface. This inconsistency was demonstrated  both
for Mode III in a square lattice and Mode I in a hexagonal
lattice, the case treated herein~\cite{mg,marderliu,kl1,pechenik}.
In Mode III, the inconsistency in the PLL is converted to a linear
instability for the finite $\alpha$ model.

\begin{figure}
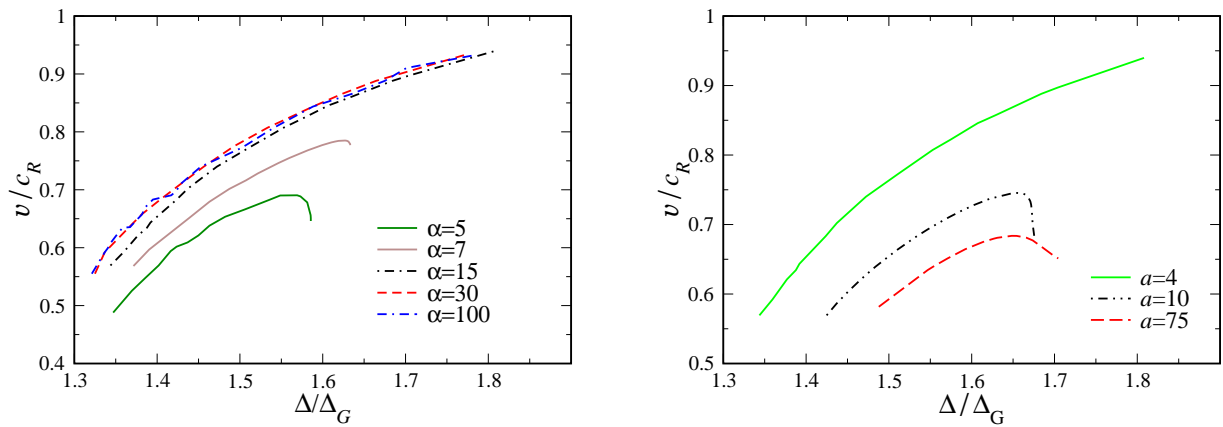

\centerline{
\includegraphics*[width=7.5cm]{6.eps}
\hskip 0.4in
\includegraphics*[width=7.5cm]{7.eps}}
\caption{(a) $v/c_{R}$ versus $\Delta/\Delta_{G}$ for $N=10$, $\eta =0.25$
and $a=4$. Data is presented for the cases $\alpha =5,\ 7,\ 15,\ 30$ and
$100$. (b) $v/c_{R}$ versus $\Delta/\Delta_{G}$ for $N=10$, $\eta =0.25$
and $\alpha =15$. Data is presented for the cases $a =4,\ 10$ and $75$.
}
\label{fig3}
\end{figure}

We present in Fig. \ref{fig3}(a) large velocity steady-state
curves for the case $N=10$ for several values of $\alpha $. The
effect of changing $\alpha$ is rather dramatic. For $\alpha=30$,
the velocity grows smoothly with $\Delta $. For $\alpha =100$ the
curve still grows but a little bit less smoothly with some small
kinks appearing. There is no sign of a maximal velocity, at least
for the range of $\Delta$ investigated; the maximal velocity, if
it exists, is close to the Rayleigh wave speed. As we go lower in
$\alpha $, at $\alpha =15$, and more significantly at $\alpha =7$
and $\alpha =5$, the curves lie significantly below those for the
larger $\alpha$. We see here a maximal velocity far below the
Rayleigh wave speed. These changes are a direct sign of the
involvement of the additional bonds in the crack progression; in
particular, the creation of a local maximum of the velocity as a
function of $\Delta$ is due to the fact that this involvement
grows with $\Delta$ and can overcompensate for the natural
tendency of the crack to speed up as the applied stress increases.

In the previous figure, the lattice scale $a$ was kept fixed. If
we increase $a$, the maximal velocity decreases and is no longer
near the Rayleigh wave speed even at larger $\alpha $. This is
presented in Fig. \ref{fig3}(b), where we show the curves for
$\alpha =15$ for different $a$.  Thus the maximum velocity is a
strongly varying function of both $\alpha$ and $a$. It will turn
out from the linear stability analysis to be presented in the next
section that there exists an instability which typically sets in
slightly after the maximal velocity point; this means that the
instability point will also be a strongly varying function of the
parameters. In the experiments, it should be noted, the critical
velocity for the onset of instabilities appears to be material
dependent\cite{review}. Thus, this strong dependence is an encouraging
sign. Also, for large $\alpha$ and $a$, the maximal
velocity is very close to the velocity at which the PLL becomes
inconsistent, as expected.

\section{STABILITY ANALYSIS}

Given the steady-state solution, we can proceed to investigate its linear
stability.  We will focus here on the instability in the high velocity
regime; it is clear that in the low velocity regime where the $v(\Delta)$ 
curve is backward, the solutions are unstable. We take the form:
\begin{equation}
\label{expand}
\vec{u}_{i}(t)=\vec{u}_{i}^{(0)}(\tau )+e^{ \frac{ \omega x}{a}}%
\vec{\delta u}_{i}(\tau )
\end{equation}
where $\tau $ is the traveling wave coordinate $\tau =t-x/v$, and $\vec{u}%
_{i}^{(0)}(\tau )$ is the steady state displacements for row $i$\
that we have found in the last chapter. We assume $\delta u\ll 1$
and expand the equations of motion to linear order. Note that
$e^{\omega x/a}=e^{\omega v(t-\tau)/a}$; thus, stability requires
$\mathop{\rm Re} (\omega )<0$, and stable perturbations decay
ahead of the crack in the moving frame of reference.

Substituting Eq. (\ref{expand}) into the equations of motion, Eq. (\ref{eom})
we get the linear equations:
\begin{equation}
\frac{d^{2}}{d\tau _{i}^{2}}\vec{\delta u}_{i}(\tau
_{i})=\sum\limits_{j\in \,nn}e^{-\omega \frac{x_{j}-x_{i}}{a}}
\left(\vec{\delta u}_j\cdot\frac{\partial}{\partial
\vec{u}_{j}^{(0)}(\tau _{j})}\right)
\left(\vec{f}_{\vec{x}_{i},\vec{x}_{j}}+\vec{g}_{\vec{x}_{i},\vec{x}_{j}}
\right)
\end{equation}
where $f$ and $g$ are the forces defined in chapter II and
$\tau _{i}=t-x_{i}/v.$

To find the possible values of $\omega $ we proceed as follows:
First we impose the boundary conditions $\vec{\delta u}_{x,y}=0.$
for the constrained rows $i=-N$, $N+1$. Also $\vec{\delta
u}_{x,y}=0$ for any $x$'s that correspond to times before or after
the limiting values of $\tau $ that were selected to build the
steady-state solution. Second, we add a normalization condition by
setting $\hat{y}\cdot \vec{\delta u}_1(0)=1.$; there are
now as many equations as variables if one includes $\omega$ among
the latter. Then, as in the steady state solution, we need to
solve a set of (now complex) linear equations with an almost
banded structure, where now $\omega $ is the variable that
destroys the banded structure. To proceed, we choose some
(complex) value of $\omega $, and temporarily ignore the actual
equation of motion for the $y$ component of $\vec{\delta
u}_{i}(0)$, replacing it by the aforementioned normalization
condition. We then solve the banded system of linear equations by
standard techniques. We still have left the equation of motion we
have dropped, which serves as a complex mismatch function; the
zeroes of this mismatch function determine the eigenvalues of
$\omega $. We are of course interested in the eigenvalue with
largest real part, which is the dominant mode.

When we solve the system of equations for the linear stability
analysis, we do not impose any symmetry on the problem; i.e. we
use the complete lattice encompassing both sides of the crack. Let
us note that any time we find a complex root of $\omega $, with
eigenfunctions $\vec{\delta u}$, there is always an equivalent solution
generated by the transformation:
\begin{eqnarray}
\mathop{\rm Im}%
\omega &\to& 2\pi -%
\mathop{\rm Im}%
\omega \nonumber \\
\vec{\delta u_{x,y}}&\to& \vec{\delta u_{x,y}}^{\ast }\ (even\,\,rows) \\
\vec{\delta u_{x,y}}&\to&-\vec{\delta u_{x,y}}^{\ast }\ (odd\,\,rows) \nonumber
\end{eqnarray}

By even and odd we mean the serial number of the row, for example
row $0$ is the first even row. This alternation is due to the
shift of $a/2$ for the $x$-values in neighboring rows.

To test our stability analysis we check whether our linear equations reproduce
the time translation mode that should be always be present with eigenvalue
$\omega =0$, and eigenfunction
\begin{equation}
\vec{\delta u}_{i}(\tau )=\frac{d}{d\tau }\vec{u}_{i}^{(0)}(\tau )\diagup \frac{d}{d\tau
}(\hat{y}\cdot\vec{u}_{1}^{(0)}(0)).
\end{equation}
Given our discrete $\delta \tau $, we found the eigenvalue of the
translation mode to lie not exactly at zero, but still quite
small; e.g., typically $\omega \sim 10^{-4}$ for $\delta \tau \sim
0.1.$ Note that compared to the case of Mode III, the matrix here
is larger as we have two components of the displacement field
Furthermore, the critical velocity here is much smaller than the
Mode III case, so $n_{b}$ is larger. It was therefore difficult to
work at $N=10$ as we did for the steady state solution. The
stability matrix is complex as opposed to real, and we do not
impose any symmetry on the problem; hence the matrix is now four
times bigger than in the steady state solution case. We chose to
work at $N=4$ to reduce memory consumption. $N=4$ is of course not
good enough to give us quantitative results for a macroscopic
system, but we expect the qualitative picture to remain the same.

\begin{figure}
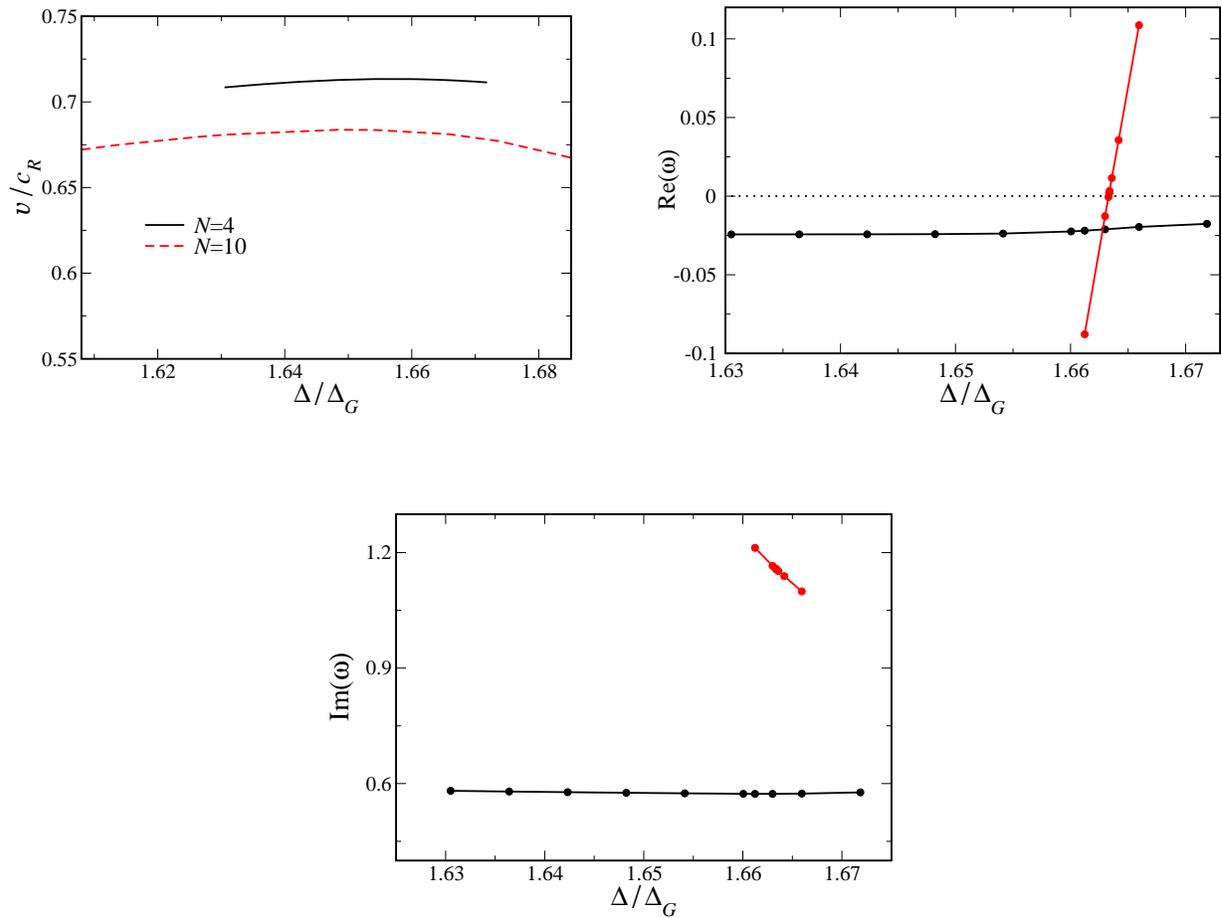

\centerline{
\includegraphics*[width=7.5cm]{8.eps}
\hskip 0.4in
\includegraphics*[width=7.5cm]{9.eps}}
\vskip 0.5in
\centerline{\includegraphics*[width=7.5cm]{10.eps}}
\caption{(a) The $v(\Delta)$ curve for the case of
$N=4,\ \eta =0.25,\ \alpha =15$ and $a=75$. We also show the
$N=10$ curve with the same parameters to demonstrate the effect of
changing $N$. (b-c) Eigenvalues for the stability problem, with
$N=4,\ \eta =0.25,\ \alpha =15$ and $a=75$. The calculation was
done with $2T+1=1000$ and $n_{b} =20$.
$\mathop{\rm Re}(\omega )$ is shown in (b),
$\mathop{\rm Im}(\omega )$ in (c).}
\label{fig4}
\end{figure}

We focus first on the case of $\alpha =15$, $a=75$, and
$\eta=0.25$. We chose a large value of $a$ because the curves are
much smoother than for small $a$.  In addition, this choice allows
for a more meaningful comparison to the results for the critical
velocity for the PLL \cite{pechenik}. In Fig \ref{fig4}(a)
we present the steady-state $v(\Delta)$ curve, along with the
corresponding curve for $N=10$.  The overall similarity of the two
curves, with an overall shift in velocity of about 5\%, lends
support to our expectation that the $N=4$ system is qualitatively
similar to those with larger $N$. In Fig. \ref{fig4}(b) we exhibit
the real parts of the two most important eigenvalues as a function
of $\Delta $, and in Fig. \ref{fig4}(c)  the imaginary parts of
the eigenvalues. We can see by looking at the steady state solution
curve and the linear stability curve that the point of instability
lies slightly after the maximum in the $v(\Delta)$ curve. This is
reasonable because both the instability and the maximum are signs
of additional bond breaking, and so should occur close together.
In the next chapter we will verify this behavior via direct
simulation. The second interesting point is that in all our
solutions there is a stable mode that almost does not change with
$\Delta $, and there is a quickly-varying mode that takes its
place as the dominant mode and crosses over to positive
$\mathop{\rm Re}(\omega )$.

\begin{figure}
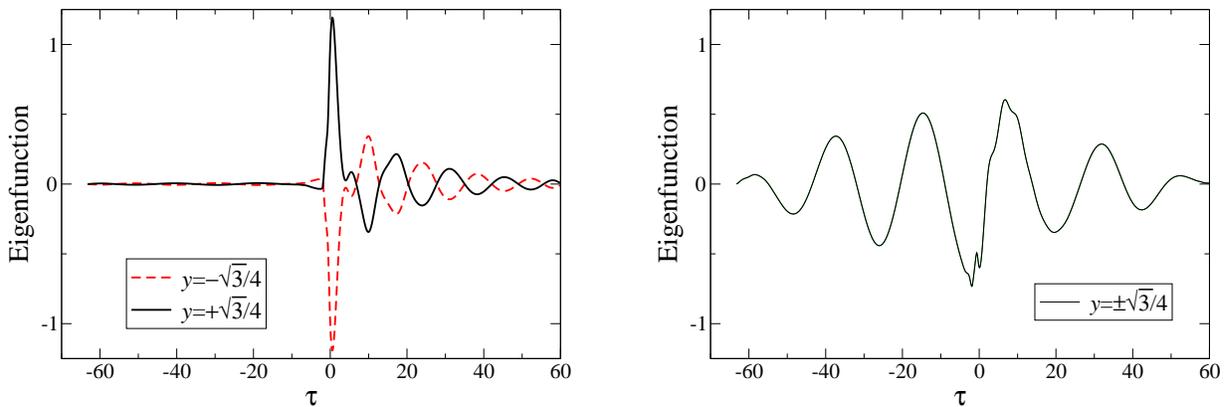

\centerline{
\includegraphics*[width=7.5cm]{11.eps}
\hskip 0.4in
\includegraphics*[width=7.5cm]{12.eps}}
\caption{Eigenfunction for stability problem, with $N=4,\ \eta =0.25,\
\alpha =15$ and $a=75$: (a) $y$ component, $\mathop{\rm Re}\hat{y}\cdot \vec{
\delta u}_{0,1}$;
 (b) $x$ component, $\mathop{\rm Re}\hat{y}\cdot \vec{\delta u}_{0,1}$.
Both figures were done with $2T+1=1000$, $n_{b} =20$ and $dt=0.13$.
}
\label{fig5}
\end{figure}

In Fig. \ref{fig5} we show a typical eigenfunction (for the
quickly-varying mode) from the linear stability analysis. Fig.
\ref{fig5}(a) presents the real part of the $y$ component of the
eigenfunction for the two rows on either side of the crack surface
($y=\pm \sqrt{3}/4$) as a function of the traveling wave
coordinate $\tau $. We see that in this case the most unstable
eigenfunction is antisymmetric about the crack surface. The
eigenfunction decays slowly downstream and very rapidly upstream
of the crack tip. In Fig. \ref{fig5}(b) we show the real part of
the $x$ component for the two rows about the midline. Here, both
upstream and downstream of the crack tip, this component of
the  eigenfunction decays
very slowly. Moreover, it is here
symmetric about the midline; the two curves overlap. We note
in passing that there is a large class of possible symmetries for
the eigenfunctions. In addition to the eigenfunction with
antisymmetric $y$ component and symmetric $x$ component, there
also exists eigenfunctions with symmetric $y$ component and
antisymmetric $x$ component. The symmetry of the imaginary part of
the eigenfunctions is also variable, and not connected in any
special way to the symmetry of the real part. Thus, while imposing
a specific symmetry reduces the size of the matrix to be solved,
it means that the calculation has to be repeated for each type of
possible symmetry.

\begin{figure}
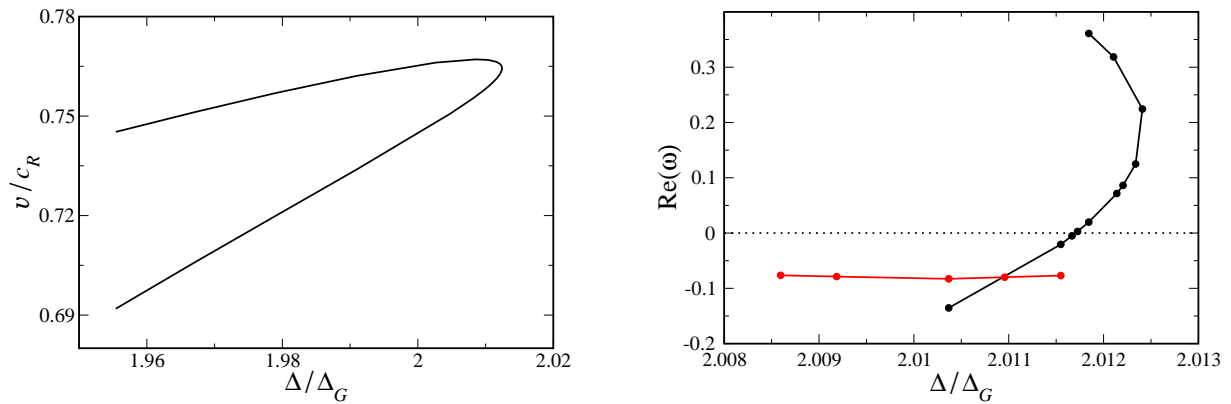

\centerline{
\includegraphics*[width=7.5cm]{13.eps}
\hskip 0.4in
\includegraphics*[width=7.5cm]{14.eps}}
\caption{(a) The $v(\Delta)$ curve for the case of $N=4,\ \eta =0.75,\
\alpha =15$ and $a=75$. (b)
$\mathop{\rm Re}(\omega )$
of eigenvalue for stability problem, with
$N=4,\ \eta =0.75,\ \alpha =15$ and $a=75$. The calculation was done with
$2T+1=1000$ and $n_{b} =20$.
}
\label{fig6}
\end{figure}

\begin{figure}
\centering{
\vskip 0.3in
\includegraphics*[width=8cm]{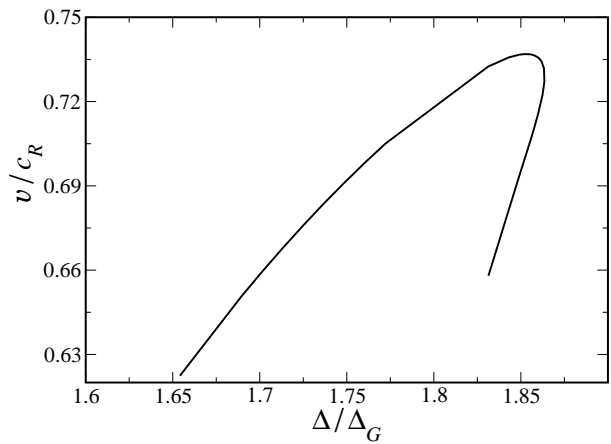}
}
\caption{The $v(\Delta)$ curve for the case of $N=4,\ \eta =0.5,\
\alpha =15$ and $a=75$.
}
\label{fig7}
\end{figure}

We next investigated a higher value of $\eta $, $\eta =0.75$. The
steady-state curve, as we see in Fig. \ref{fig6}(a), rises to a
maximum and then it comes back around in a lower branch. In the
linear stability analysis for this $\eta $, Fig. \ref{fig6}(b),
we see that the point of instability is exactly at the $\Delta
$-maximum; all the points on the backward branch are unstable.
This is of course a generic feature of propagating systems. At the
$\Delta$ maximum, the velocity of the solution is indeterminate to
linear order, implying a zero mode at this point. This zero mode
represents the crossing over into instability of some mode of the
system. When we decrease $\eta $ the backward branch comes back
less rapidly, as shown in Fig. \ref{fig7} 
for $\eta =0.5$. Clearly at some smaller $\eta$ this
turnaround vanishes altogether, and we recover the situation at
$\eta =0.25$, where no maximum $\Delta$ is seen. The linear
stability analysis for the $\eta =0.5$ reveals that the maximal
$\Delta$ is no longer the location of the first instability; there
is another mode which goes unstable before this point.

\begin{figure}
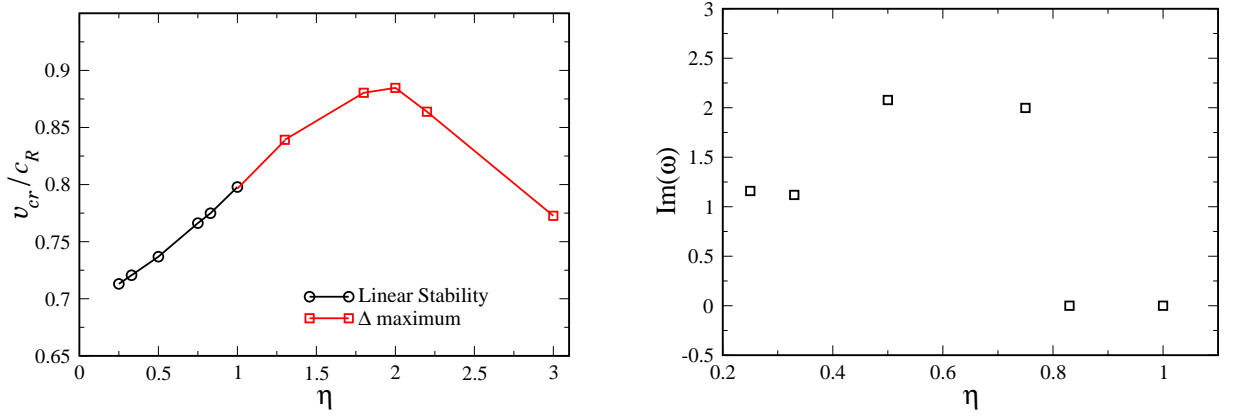

\centerline{
\includegraphics*[width=7.5cm]{16.eps}
\hskip 0.4in
\includegraphics*[width=7.5cm]{22.eps}}
\caption{(a) The critical velocity from the linear stability analysis,
normalized
by the Rayleigh wave speed, as a function of $\eta$. Data is presented for
$\alpha =15$, $a=75$ and $N=4$. (b) $\mathop{\rm Im}(\omega )$ of the
eigenvalue at the critical velocity as a function of $\eta$.
}
\label{fig8}
\end{figure}

We present in Fig. \ref{fig8}(a) a curve of the critical velocity
for the onset of the instability as a function of $\eta$. Given
the correlation we found between the critical velocity for
instability and the point of maximal $\Delta$ for sufficiently
large $\eta$, we extended the graph to higher $\eta$ by
calculating the velocity at the $\Delta$ maximum. The critical
velocity for small $\eta $ increases with $\eta $, as in the
piece-wise linear model (where here the critical velocity denotes
the onset of inconsistency); even the values are similar. As we
increase $\eta$ further, the critical velocity decreases, again
just like the piece-wise linear results, due there to the breaking
of a horizontal bond~\cite{pechenik}. We further see that the
critical velocity extrapolates to a finite value at $\eta =0$.
Thus, there is stable propagation below this critical velocity at
zero dissipation in contradiction to the claim of Pla, et
al.\cite{Sander}. We believe that the instabilities they saw in
their simulations were finite amplitude effects engendered by the
initial conditions. Indeed, we have generated stable propagating
solutions at $\eta =0$ via direct simulation (see next section).

Note that in the Mode I case as opposed to the Mode III
case~\cite{nonlinear}, $\mathop{\rm Im} (\omega )$ at the point of
instability is not close to $\pi $. Therefore, there is no
parallel to the period doubled nonlinear state found in that
system. The critical $\mathop{\rm Im}(\omega )$ here varies strongly with
$\eta $. We present in Fig. \ref{fig8}(b) the $\mathop{\rm
Im}(\omega )$  at the point of instability as a function of
$\eta$. We see two transitions between dominant modes; somewhere
in the range  $0;.78 <\eta< 0.83$, the critical $\mathop{\rm Im}(\omega )$
becomes zero, as it must if the transition is directly connected to
the saddle-node behavior of the steady-state curve. At lower $\eta $, e.g., $\eta =0.5$, it was about $%
2.0$, as mentioned above. Note that this transition is sharp;
for example at $\eta=0.83$ we still have the mode with $\mathop{\rm Im}(\omega
)\simeq 2$ but the $\mathop{\rm Im}(\omega)=0$ mode goes unstable first.
We also have a transition between dominant modes at $%
\eta \simeq 0.4$, the modes having very different $\mathop{\rm
Im}(\omega)$. At higher $\eta$ we checked the critical
$\mathop{\rm Im}(\omega)$ by the direct simulation (again, see the
next section) and found that it remains zero.

\section{COMPARISON TO SIMULATION}

To check our results, both for the steady state solution and for
the linear stability analysis we have implemented a direct
numerical simulation of the initial value problem. These
simulations are also useful for investigating the dynamics after
the point of instability. Our system is a hexagonal lattice with
$2N+1$ rows and with a finite number, $L$, of mass points in each
row. The top and bottom rows are constrained to have a
displacement $\pm\Delta\hat{y} $. The initial conditions for the
displacement and velocity for the lattice were constructed from
the steady state solution that we already have found, with the
initial crack tip placed $L/3$ from the left edge. We solve the
equations of motion by an Euler scheme, taking
$\vec{v}_{t+dt}=\vec{v}_{t}+
\mathop{\displaystyle\sum}%
\limits_{nn}(\vec{f}+\vec{g})dt$ and $\vec{x}_{t+dt}
=\vec{x}_{t}+\vec{v}_{t+dt}dt$. We measured the velocity of the
crack by monitoring the bond extension. The crack was considered
to have translated forward when both bonds connecting a given mass
point to its neighbors across the midline exceeded the critical
extension, $\varepsilon= 1$, and the velocity determined from the
time $\Delta t$ between such translation events, $v=a/\Delta t$.
This criterion of velocity is of course somewhat arbitrary because
cracking in our smooth potential is a reversible process. Note
that to produce a true steady state crack we need to fix $a/v$ so
that it is an integer multiple of $dt$. Otherwise we would
introduce oscillations in $\Delta t$ due to the
incommensurability. Via this procedure, we obtained excellent
agreement (with 4 digit accuracy) between the simulated and
calculated results for the $v(\Delta) $ relationship. As we
increase $\Delta $, we can follow the $v(\Delta) $ curve and we
obtain from the simulation all the points of the curve including
those around the $v$-maximum. When we get exactly to the point of
instability as predicted from the linear stability analysis,
additional bond breaking occurs, and the velocity in the
simulation drops down below the steady-state curve.

\begin{figure}
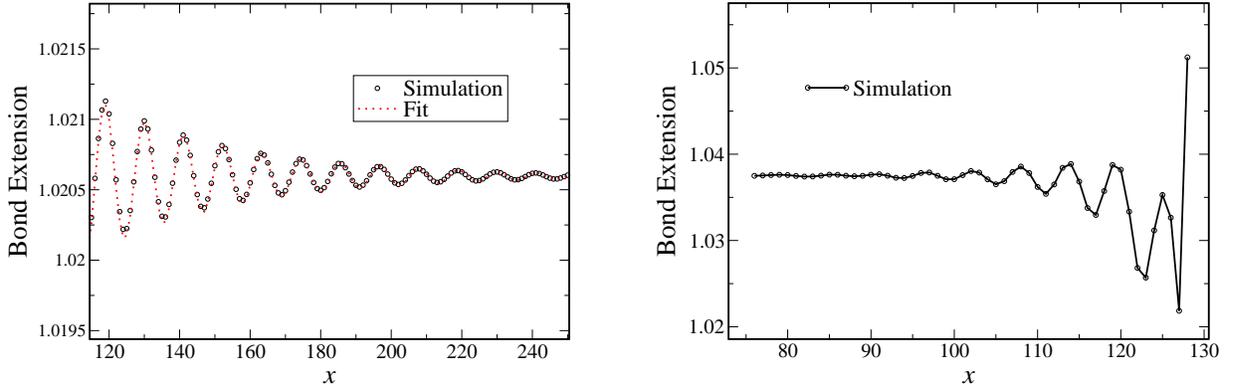

\centerline{
\includegraphics*[width=7.5cm]{17.eps}
\hskip 0.4in
\includegraphics*[width=7.5cm]{18.eps}
}
\caption{(a) Bond extension at breaking as a function of position $x$ along the
crack surface. Here $N=4,\ \eta=0.25,\ \alpha=15$ and $a=75$. $\Delta/\Delta
_{G}=1.6305, \ dt=0.125$. The fitted curve is $1.021+0.01e^{-0.025x}\sin(0.565x
+3.5)$. (b) Bond extension at breaking as a function of position $x$ after the
point of instability for $N=4,\ \eta=0.25,\ \alpha=15$ and $a=75$.
}
\label{fig9}
\end{figure}

The numerical simulation can provide a strong check on the details
of the stability analysis. We can measure the eigenfrequency by
perturbing one mass point on the lattice in the vicinity of the
crack tip, and measuring the subsequent time development of the
perturbation. We track the bond extension of the breaking bond at
the moment of cracking. Because of the finiteness of $dt$ this is
always somewhat larger than $\varepsilon$. This value oscillates,
decaying in the stable regime, and growing in the unstable regime.
The data can then be fit to a form $Ae^{\frac{ \omega x}{a}}$ and
$\omega$ compared to that produced by the linear stability
analysis. In Fig. \ref{fig9}(a) we can see the results for the
crack tip and if we perform the fitting, the eigenvalue found here
lies quite close to the eigenvalue found in the linear stability
analysis, $\omega
=-0.0243+0.5813i$.
If we instead look at a point in the unstable regime, this
oscillations look like those presented in Fig. \ref{fig9}(b). The
oscillating part in the simulation fits well to the imaginary part
of the linear stability eigenvalue. We note that throughout the
unstable regime we can detect the presence of the square of the unstable
mode, giving rise to an additional non-linear mode whose
real part is two times the value of the real part of the
eigenvalue.

\begin{figure}
\centering{ (a)\includegraphics*[width=8.5cm]{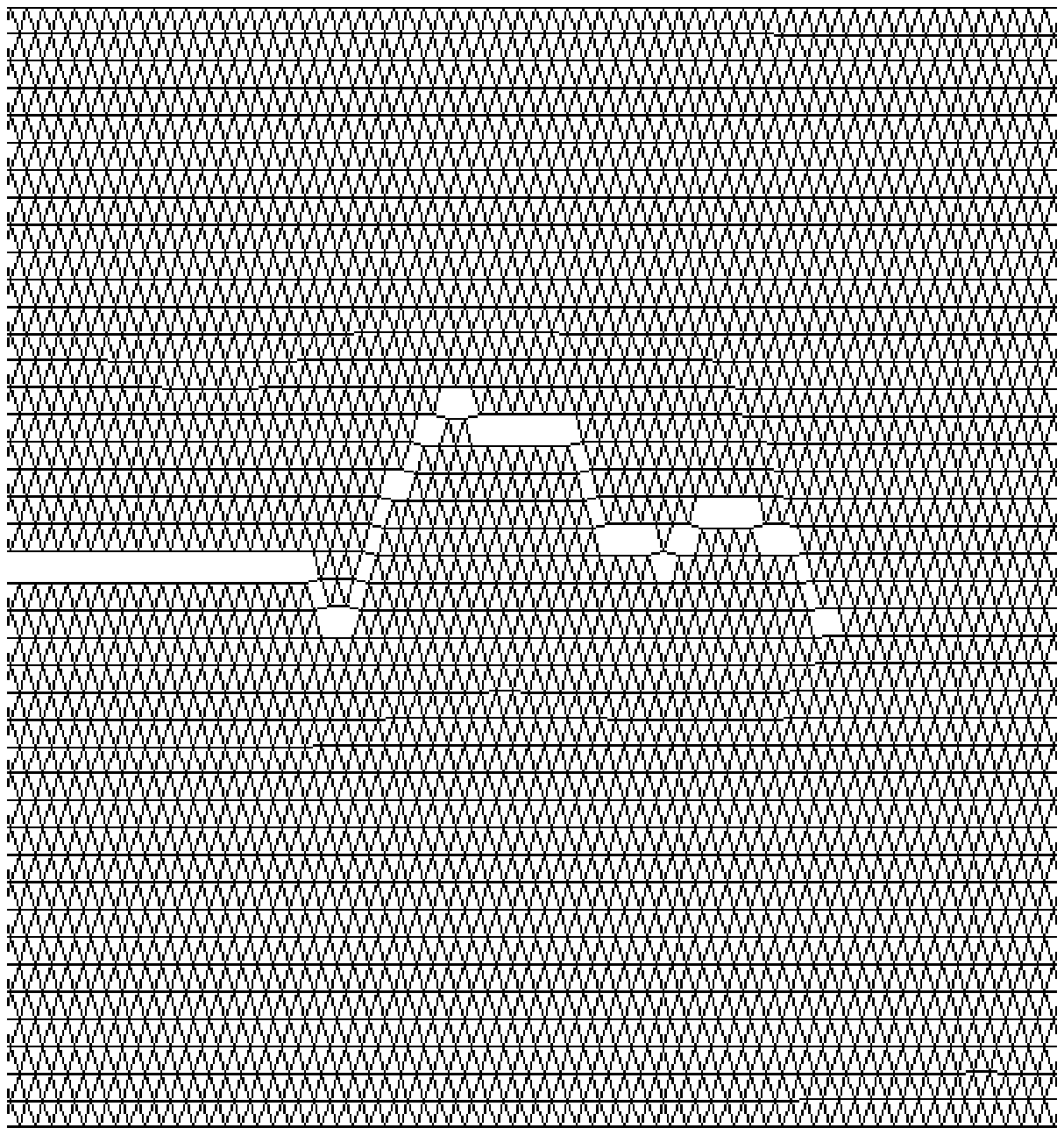}\\
(b)\includegraphics*[width=6.0cm]{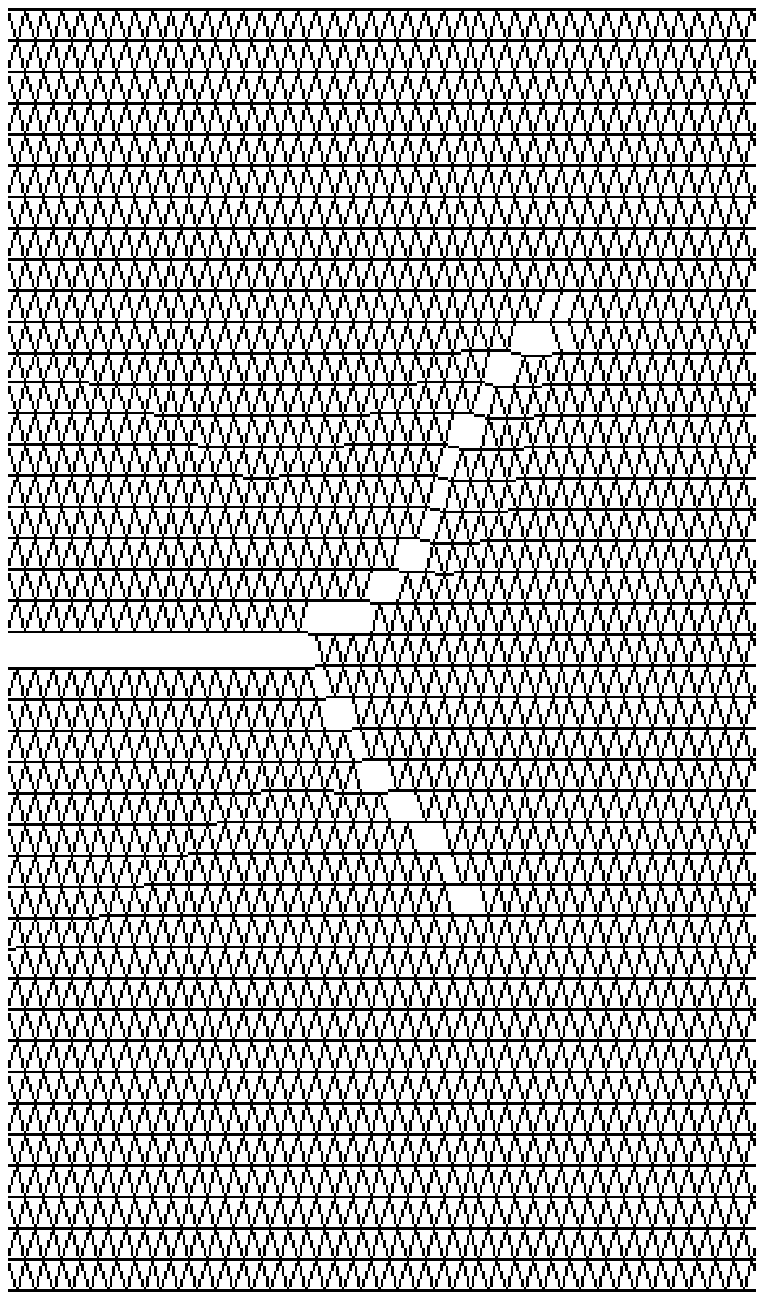}
(c)\includegraphics*[width=5.7cm]{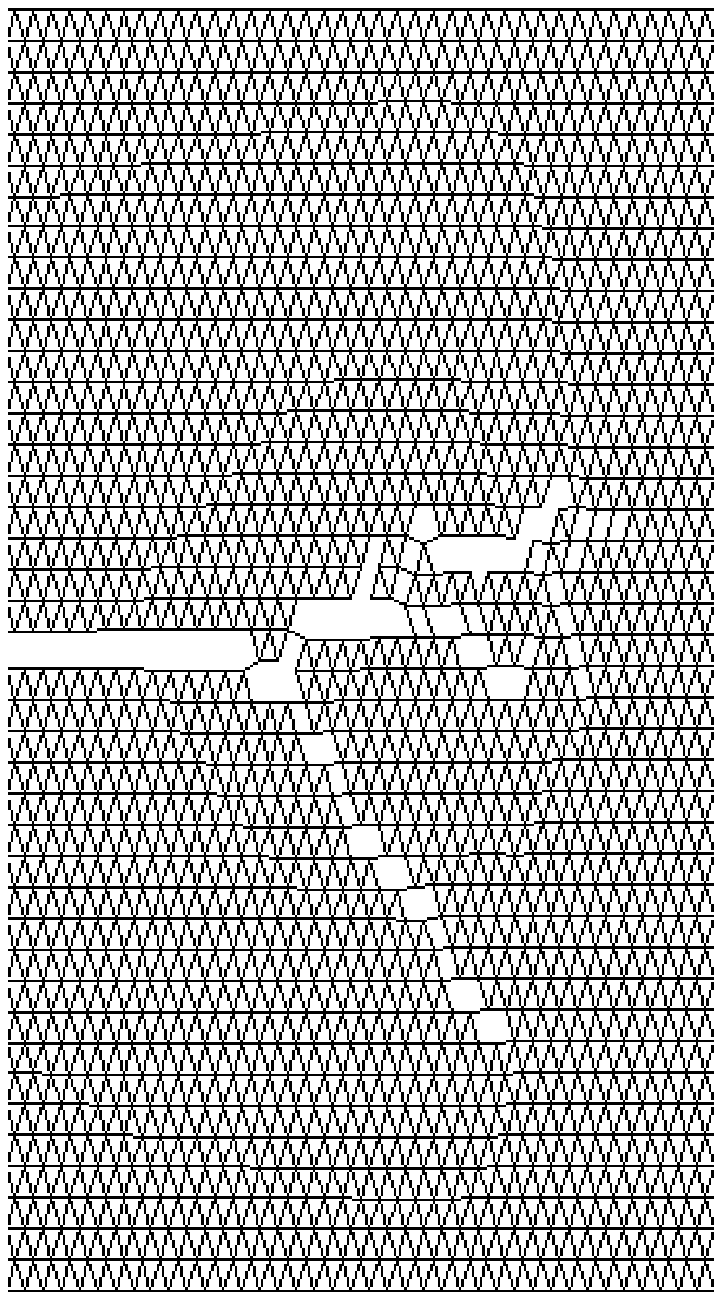} } \caption{(a) The lattice
showing the broken bonds for $N=20, \ \eta =0.1, \ \alpha =15$, $a=75$.
$\Delta/\Delta_{G}=1.875$. (b) $\Delta/\Delta_{G}=2.1875$. (c)
$\Delta/\Delta_{G}=3.125$. } \label{fig10}
\end{figure}

\begin{figure}
\centering{
\vskip 0.3in
\includegraphics*[width=10cm]{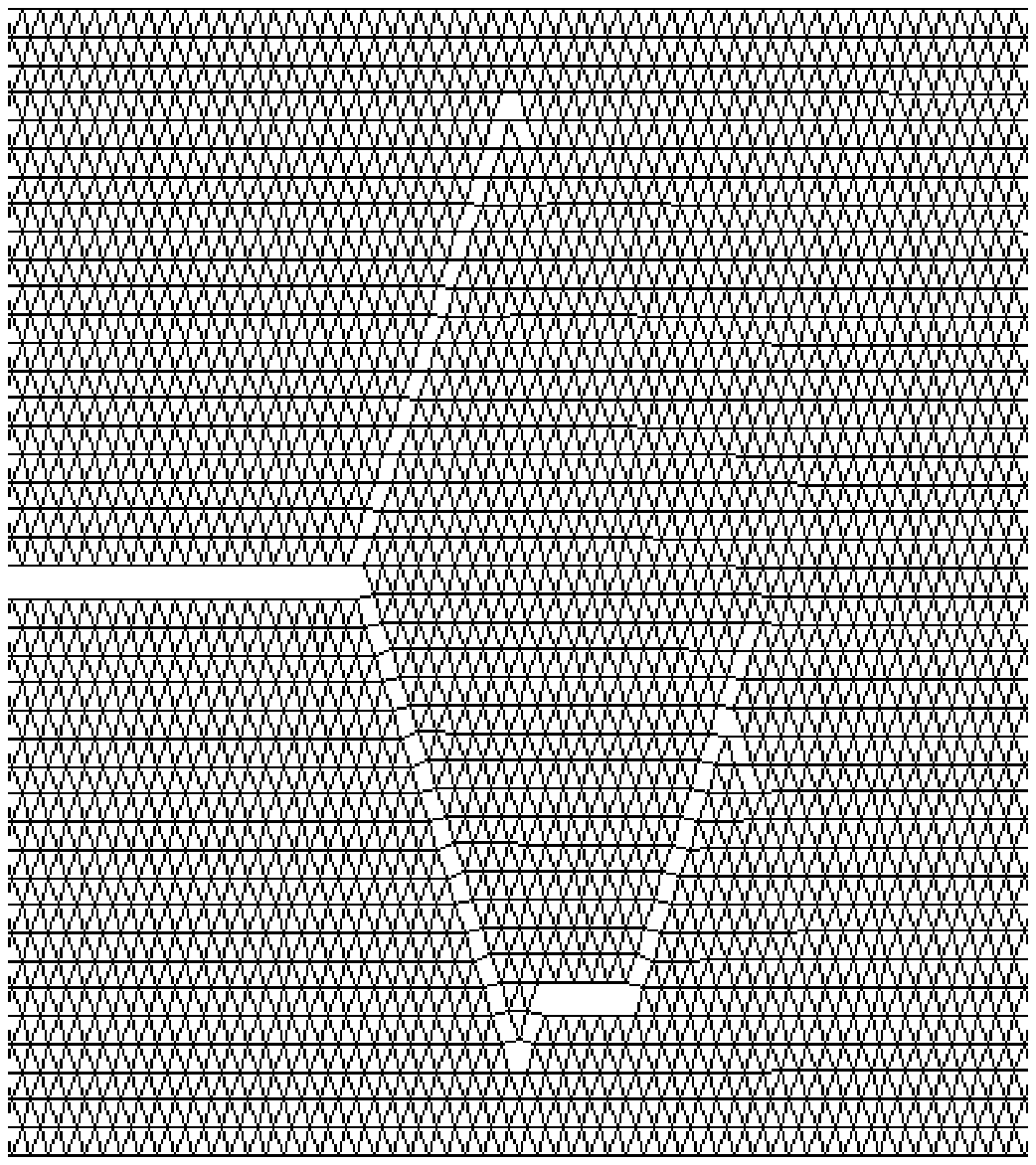}}
\caption{The lattice showing the broken bonds for $N=20, \ \eta =1.2, \ \alpha =15$, $a=75$.
$\Delta/\Delta_{G}=2.578$.}
\label{fig11}
\end{figure}

When we increase $\Delta $ past the point of instability, the
actual crack dynamics of the problem become increasingly complex.
Direct numerical simulation an still be used to study crack
development. We performed such a study using $N=20$. We see in
Fig. \ref{fig10} three examples of what can happen in Mode I
cracking after the point of instability, for increasing values of
$\Delta$, all for the case of small dissipation. For the smallest
$\Delta$, the crack tip meanders up and down off the midline. For
larger $\Delta$, the crack starts to bifurcate (Fig.
\ref{fig10}(b)), and at the highest $\Delta$ shown bifurcates many
times and creates a very complex cracking picture (Fig.
\ref{fig10}(c)). In all three of these small $\eta$ cases, the
crack arrests eventually. At large $\eta$, such as $\eta =1.2$ and
higher (see Fig. \ref{fig11}) the picture is very different. Just
above the point of instability the crack bifurcates, with the two daughter
tips propagating close to the edges of the system.  These kinds of behavior
appear to be very different from that seen in experiment, where there is
a main crack propagating more or less straight in addition to daughter 
cracks on either side.

In the Mode III case, a crucial point that emerged from the
detailed linear stability study was that the additional cracking
bonds were always created behind the crack tip. There, the
instability did not divert the main crack, but merely slowed it
down temporarily as energy was diverted to create the
sidebranches~\cite{nonlinear,kkl}. 
Here in Mode I cracking, however, after the point of instability
the crack tip itself changes direction and the crack either bifurcates or
deviates from the center.


\section{SUMMARY AND DISCUSSION}

We have herein studied the steady-state crack, calculating
 the $v(\Delta)$ curves both for small and large
driving, and for small and large dissipation. For small driving
with small dissipation, we obtained oscillating curves for all
finite $\alpha$, indicating a substantial velocity gap at small
velocity. At large dissipation, the curves are smooth, and vary
significantly with $\alpha$. At large driving, the maximal
velocity exhibits a strong dependence on the smoothness and the
value of $a/\varepsilon$. For large $\eta$, there is also a
maximal driving as the curve bends back forming a second branch,
and beyond which no steady-state solution exists. For all $\eta$,
the steady-state solution undergoes a linear instability. For
large $\eta$, this occurs at the maximal driving, and is a real
instability. For small $\eta$, the instability set in slightly
after the maximal velocity, and is a Hopf bifurcation. The
velocity at the onset of instability initially increases with $\eta$, reaches
a maximum at $\eta \simeq 2$ and thereafter decreases for larger $\eta$. We
note that the strong variation of the velocity at the onset of
instability with the microscopic parameters does not accord with
the naive expectations based on the Yoffe continuum calculation\cite{yoffe} of
a change in the direction of maximal stress at some universal
velocity, independent of the microscopic dynamics. 
It also is not accord with the ansatz of
Eshelby based on energy considerations~\cite{eshelby}. We also
compared our steady state solution and linear stability analysis
to direct numerical simulations, obtaining excellent agreement.

We plan to extend our work to the case of biaxial loading, where the
material in strained in both the $x$ and $y$ direction. This
phenomena describes for example a popping of a balloon as
described in Ref.~\cite{balloon}. We also plan to investigate
further the post-instability dynamics. Finally, we hope to use the
recently introduced~\cite{phase} continuum regularization of
tip-dynamics (based on the phase-field method) to unravel the role
that the lattice structure has in determining the form of the
allowed instabilities.

\begin{acknowledgments}
The work of DAK is supported in part by the Israel Science
Foundation. HL is
supported in part by the US NSF under grant DMR98-5735.
\end{acknowledgments}

\end{document}